\documentclass[12pt]{article}
\usepackage{a4wide,amsmath,amssymb,graphicx}
\numberwithin{equation}{section}
\DeclareMathOperator{\Tr}{Tr}
\newcommand{\D}{\rlap{\hspace{0.2em}/}D}

\title{Asymptotic freedom:\\
history and interpretation%
\thanks{Lecture given at VI Dubna International Advanced School
of Theoretical Physics, Dubna, 2008}}

\author{Andrey Grozin\\
Institut f\"ur Theoretische Teilchenphysik,
Universit\"at Karlsruhe}

\date{}

\begin{document}

\maketitle

\begin{flushright}
\begin{picture}(0,0)
\put(0,60){\makebox(0,0)[r]{\large TTP08-12}}
\end{picture}
\end{flushright}

\begin{abstract}
In this lecture, the early history of asymptotic freedom is discussed.
The first completely correct derivation of $\beta_0$
in non-abelian gauge theory (Khriplovich, 1969) was done in the Coulomb gauge;
this derivation is reproduced (in modernized terms) in Sect.~\ref{S:Khr}.
A qualitative physical explanation of asymptotic freedom
via chromomagnetic properties of vacuum (Nielsen, 1981)
is discussed in Sect.~\ref{S:Niels}.
\end{abstract}

\section{Introduction}
\label{S:Intro}

Evolution of the QED coupling $\alpha(\mu)$ is determined
by the renormalization group equation
\begin{equation}
\mu \frac{d\alpha(\mu)}{d\mu} =
- 2 \beta(\alpha(\mu))\,\alpha(\mu)\,,\qquad
\beta(\alpha) = \beta_0 \frac{\alpha}{4\pi}
+ \beta_1 \left(\frac{\alpha}{4\pi}\right)^2
+ \cdots
\label{Intro:RG}
\end{equation}
where the 1-loop $\beta$-function coefficient is
\begin{equation}
\beta_0 = - \frac{4}{3}
\label{Intro:beta0}
\end{equation}
(we use the $\overline{\text{MS}}$ renormalization scheme).
The sign $\beta_0<0$ corresponds to the natural picture
of charge screening --- when the distance grows ($\mu$ reduces),
$\alpha(\mu)$ becomes smaller.

Non-abelian gauge theories have $\beta_0>0$.
This corresponds to the opposite behaviour
called \emph{asymptotic freedom} ---
the coupling becomes small at large $\mu$ (small distances).
Discovery of this fact came as a great surprise,
and completely changed ideas about applicability
of quantum field theory for describing the nature.
In fact, it was done several times, by several researchers,
independently.

The first paper in which $\beta_0>0$ was obtained
is by Vanyashin, Terentev~\cite{VT:65}.
They considered non-abelian gauge theory
with the gauge boson mass term introduced by hand.
Of course, this theory is non-renormalizable.
It contains longitudinal gauge bosons
whose interaction grows with energy.
They obtained
\begin{equation}
\beta_0 = \left(\frac{11}{3} - \displaystyle\frac{1}{6} \right) C_A\,.
\label{Intro:VT}
\end{equation}
The contribution $-1/6$ comes from longitudinal gauge bosons.
In the massless gauge theory it is canceled by the ghost loop
(which was not known in 1965).
Now we know that renormalizable massive gauge theory
can be constructed using the Higgs mechanism.
In this theory, the result~(\ref{Intro:VT}) is correct;
the contribution  $-1/6$ comes from the Higgs loop.

The first completely correct derivation of $\beta_0$
in non-abelian gauge theory has been published by Khriplovich~\cite{Kh:69}.
He used the Coulomb gauge which is ghost-free.
The Ward identities in this gauge are simple (as in QED),
and it is sufficient to renormalize the gluon propagator
in order to renormalize the charge.
This derivation clearly shows why screening is the natural behaviour,
and how non-abelian gauge theories manage to violate
these general arguments.
It is also very simple.
We shall discuss this derivation of $\beta_0$ in Sect.~\ref{S:Khr},
following~\cite{Kh:69}, but using a modernized language.

't~Hooft discovered asymptotic freedom in 1971--72
while studying renormalization of various field theories
in dimensional regularization~\cite{tH:72}.
He reported his result during a question session
after Symanzik's talk at a small conference in Marseilles in June 1972.

And finally, Gross, Wilczek~\cite{GW:73} and Politzer~\cite{P:73}
discovered it again in 1973.
They were first who suggested to apply asymptotically free gauge theory (QCD)
to strong interactions, in particular, to deep inelastic scattering.

The early history of asymptotic freedom is discussed in several papers,
e.g., \cite{Sh:01,tH:01,W:96},
and in the Nobel lectures~\cite{Nobel}.

Now $\beta_0$ is derived in every quantum field theory textbook.
In the standard approach, the covariant gauge is used
(with Faddeev--Popov ghosts, of course).
The coupling constant renormalization can be obtained
from renormalizing any vertex in the theory
together with all propagators attached to this vertex.
Usually, the light-quark -- gluon vertex or the ghost -- gluon vertex is used
(calculations are slightly shorter in the later case).
The 3-gluon vertex or even the 4-gluon one~\cite{W2}
can also be used.
All these standard derivations can be found, e.g., in~\cite{G:07}.
The infinitely-heavy-quark -- gluon vertex can also be used~\cite{G:04};
this derivation is, perhaps, as easy as the ghost -- gluon one.

If the Ward identities are QED-like, it is sufficient to renormalize
the gluon propagator; no vertex calculations are required.
One such case is the Coulomb gauge~\cite{Kh:69};
it has a slight disadvantage of being not Lorentz invariant
(this makes gluon loop calculations more difficult).
Another example is the background field formalism~\cite{A:81}.
It is Lorentz invariant, and provides, probably,
the shortest derivation of $\beta_0$ in QCD.

However, all these derivations don't explain asymptotic freedom (antiscreening)
in simple physical terms.
Probably, the simplest explanation of this kind is presented in~\cite{N:81}
(it is also discussed in~\cite{W:96}).
We shall discuss in in Sect.~\ref{S:Niels}.

\section{Coulomb gauge (Khriplovich 1969)}
\label{S:Khr}

\subsection{Feynman rules}
\label{S:Feyn}

This gauge has several advantages.
It is ghost-free, and Ward identities in it are simple.
It also has one significant disadvantage:
it is not Lorentz-invariant.
Currently, it is widely used in non-relativistic QCD~\cite{NRQCD}.
In this effective theory,
there is no Lorentz invariance from the very beginning,
and the disadvantage does not matter.
On the other hand,
separating Coulomb and transverse (chromomagnetic) gluons
helps to estimate various contributions
when considering non-relativistic bound states.

The gluon propagators in this gauge are
\begin{equation}
\raisebox{-0.2mm}{\includegraphics{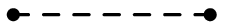}}
= - i \delta^{ab} D_0(q)\,,\qquad
\raisebox{-0.2mm}{\includegraphics{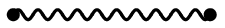}}
= - i \delta^{ab} D_0^{ij}(q)\,.
\end{equation}
The Coulomb gluon propagator
\begin{equation}
D_0(q) = - \frac{1}{\vec{q}\,^2}
\label{GlueProp:Coulomb}
\end{equation}
does not depend on $q_0$.
This means that the Coulomb gluon propagates instantaneously.
The transverse gluon propagator
\begin{equation}
D_0^{ij}(q) = \frac{1}{q^2+i0}
\left(\delta^{ij}-\frac{q^i q^j}{\vec{q}\,^2}\right)
\label{GlueProp:Transverse}
\end{equation}
describes a normal propagating massless particle
(the usual denominator $q^2+i0$).
In accordance with the gauge-fixing condition
$\vec{\nabla}\cdot\vec{A}^a=0$,
this 3-dimensional tensor is transverse to $\vec{q}$.

The three-gluon vertices look as usual.
There is no vertex with three Coulomb legs
(if we contract the usual three-gluon vertex
with $v^\mu=(1,\vec{0})$ in all three indices,
it vanishes).

We are going to calculate the potential
between an infinitely heavy quark and an infinitely heavy antiquark
in the colour-singlet state.

The infinitely-heavy quark propagator
\begin{equation}
\raisebox{-10.2mm}{\includegraphics{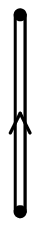}}
= \frac{i}{p_0+i0}
\label{QuarkProp}
\end{equation}
does not depend on $\vec{p}$
(if the mass $M$ were large but finite, it would be
\begin{equation*}
\frac{i}{p_0 - \frac{\vec{p}\,^2}{2M} + i0}\,;
\end{equation*}
the kinetic energy disappears at $M\to\infty$).
This means that the coordinate-space propagator is
$\sim\delta(\vec{r}\,)\theta(t)$:
the infinitely heavy quark does not move in space,
and propagates forward in time.
The energy of an on-shell infinitely heavy quark is 0,
and does not depend on $\vec{p}$
(it would be $\vec{p}\,^2/(2M)$
if $M$ were large but finite).

An infinitely heavy quark interacts with Coulomb gluons
but not with transverse ones:
\begin{equation}
\raisebox{-10.2mm}{\includegraphics{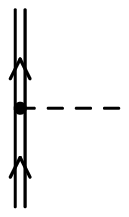}}
=i g_0 t^a\,,\qquad
\raisebox{-10.2mm}{\includegraphics{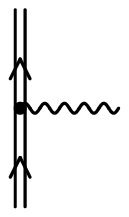}}
= 0\,.
\label{QuarkVert}
\end{equation}
The infinitely-heavy antiquark -- Coulomb gluon vertex
differs from~(\ref{QuarkVert}) by a minus sign.

\subsection{Quark--antiquark potential}
\label{S:Pot}

\begin{figure}[ht]
\begin{center}
\begin{picture}(20,36)
\put(11,18){\makebox(0,0){\includegraphics{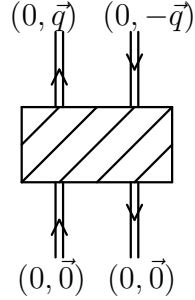}}}
\put(5,3){\makebox(0,0)[t]{$(0,\vec{0})$}}
\put(17,3){\makebox(0,0)[t]{$(0,\vec{0})$}}
\put(4,33){\makebox(0,0)[b]{$(0,\vec{q})$}}
\put(18,33){\makebox(0,0)[b]{$(0,-\vec{q})$}}
\end{picture}
\end{center}
\caption{On-shell scattering amplitude}
\label{F:QQbarScat}
\end{figure}

We calculate the scattering amplitude
of an on-shell infinitely heavy quark
and an on-shell infinitely heavy antiquark
in the colour-singlet state
with the momentum transfer $\vec{q}$
in quantum field theory (Fig.~\ref{F:QQbarScat}),
and equate it to the same amplitude
in quantum mechanics.
In the Born approximation, it is
\begin{equation}
i U_{\vec{q}}\,.
\label{SQM}
\end{equation}
Two-particle-reducible diagrams in quantum field theory
(which can be cut into two disconnected parts
by cutting a quark line and an antiquark one)
correspond to higher Born approximations in quantum mechanics,
and we don't have to consider them.

To the lowest order in $g_0^2$, the scattering amplitude is
\begin{equation}
\raisebox{-10.2mm}%
{\begin{picture}(19,22)
\put(9.5,11){\makebox(0,0){\includegraphics{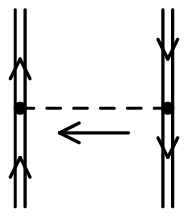}}}
\put(9.5,5){\makebox(0,0){$(0,\vec{q})$}}
\end{picture}}
= - i C_F g_0^2 D(0,\vec{q})
= i C_F \frac{g_0^2}{\vec{q}\,^2}\,.
\label{SQFT}
\end{equation}
Therefore,
\begin{equation}
U_{\vec{q}} = C_F g_0^2 D(0,\vec{q})
= - C_F \frac{g_0^2}{\vec{q}\,^2}\,,
\label{SQFT2}
\end{equation}
and we obtain the Coulomb attraction potential
\begin{equation}
U(r) = - C_F \frac{\alpha_s}{r}\,.
\label{Ur}
\end{equation}

\begin{figure}[ht]
\begin{center}
\begin{picture}(20,33)
\put(10,16.5){\makebox(0,0){\includegraphics{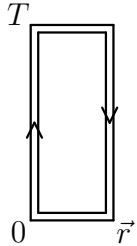}}}
\put(3,2){\makebox(0,0){$0$}}
\put(17,2){\makebox(0,0){$\vec{r}$}}
\put(3,31){\makebox(0,0){$T$}}
\end{picture}
\end{center}
\caption{Wilson loop}
\label{F:Wilson}
\end{figure}

Some people prefer to discuss
the quark--antiquark potential
in terms of the vacuum average
of a Wilson loop (Fig.~\ref{F:Wilson})
with $T\gg r$.
Of course, this is exactly the same thing,
because the infinitely-heavy quark propagator
\emph{is} a straight Wilson line
along the 4-velocity $v$.

The energy of a colour-singlet quark--antiquark pair
separated by $\vec{r}$ is $U(\vec{r}\,)$.
Therefore, neglecting boundary effects near time $0$ and $T$,
the Wilson loop is
\begin{equation}
e^{-i\,U(\vec{r}\,)\,T}\,,
\label{Wilson}
\end{equation}
or, to the first order in $\alpha_s$,
\begin{equation*}
1 - i\,U(\vec{r}\,)\,T\,.
\end{equation*}
This order-$\alpha_s$ term is
\begin{equation}
\raisebox{-15.7mm}{\begin{picture}(24,33)
\put(10,16.5){\makebox(0,0){\includegraphics{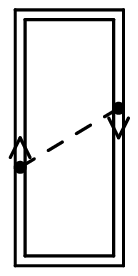}}}
\put(3,2){\makebox(0,0){$0$}}
\put(17,2){\makebox(0,0){$\vec{r}$}}
\put(3,31){\makebox(0,0){$T$}}
\put(2.5,13.5){\makebox(0,0){$\tau$}}
\put(21.5,19.5){\makebox(0,0){$\tau+t$}}
\end{picture}}
= - i\,C_F\,g_0^2\,T\,\int D(t,\vec{r})\,dt
= - i\,C_F\,g_0^2\,T\,\int \frac{d^{d-1}\vec{q}}{(2\pi)^{d-1}}\,
D(0,\vec{q})\,e^{i\,\vec{q}\vec{r}}
\label{Wilson1}
\end{equation}
(integration in $\tau$ gives $T$),
and hence we obtain~(\ref{Ur}).

\subsection{Corrections to the scattering amplitude}
\label{S:Corr}

Now we want to calculate the first correction
to the scattering amplitude in Fig.~\ref{F:QQbarScat}.

First of all, there are external-leg renormalization factors.
They are given by the derivative of the infinitely-heavy quark
self-energy at its mass shell.
The infinitely heavy quark only interacts with Coulomb gluons,
and its self-energy is
\begin{equation}
\raisebox{-15.2mm}{\includegraphics{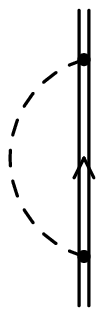}}
\sim \int \frac{d^{d-1}\vec{k}}{\vec{k}\,^2} = 0
\label{Sigma}
\end{equation}
(because the infinitely-heavy quark propagator
does not depend on $\vec{k}$).
In the coordinate space,
the infinitely heavy quark propagates along time,
and the Coulomb gluon --- along space.
Therefore, the two vertices are at the same space--time point,
and the self-energy is
\begin{equation}
\sim D(t=0,\vec{r}=0)
\sim \int \left.\frac{d^{d-1}\vec{k}}{\vec{k}\,^2} e^{i\vec{q}\cdot\vec{r}}
\right|_{\vec{r}=0} \sim U(\vec{r}=0) \Rightarrow 0\,.
\label{SigmaC}
\end{equation}
This is the classical self-energy of a point charge,
and it is linearly divergent.
In dimensional regularization, it is $0$%
\footnote{This linear ultraviolet divergence leads to
an ultraviolet renormalon singularity at Borel parameter $u=1/2$,
and hence to an ambiguity of the on-shell heavy-quark mass
proportional to $\Lambda_{\text{QCD}}$~\cite{BB:94}
(see also~\cite{G:04}, Chapter~8).}.

Transverse gluons don't interact with the infinitely heavy quark,
and hence there is only one vertex correction.
It vanishes for the same reason:
\begin{equation}
\raisebox{-15.2mm}{\includegraphics{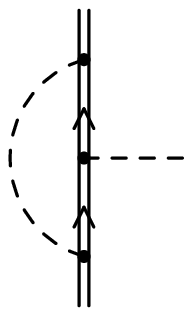}}
= 0\,.
\label{Gamma}
\end{equation}

\begin{figure}[ht]
\begin{center}
\begin{picture}(24,22)
\put(12,11){\makebox(0,0){\includegraphics{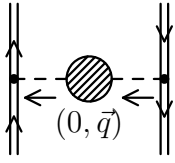}}}
\put(12,5){\makebox(0,0){$(0,\vec{q})$}}
\end{picture}
\end{center}
\caption{Vacuum polarization corrections}
\label{F:VacPol}
\end{figure}

Therefore, we only need to consider vacuum polarization corrections
(Fig.~\ref{F:VacPol}):
\begin{equation}
U_{\vec{q}} = C_F g_0^2 D(0,\vec{q})\,,
\label{Uq}
\end{equation}
where the Coulomb-gluon self-energy is
\begin{equation}
\raisebox{-3.2mm}{\includegraphics{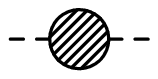}}
= - i \vec{q}\,^2 \Pi(q)\,,
\label{Pi}
\end{equation}
and its propagator is
\begin{equation}
D(q) = - \frac{1}{\vec{q}\,^2} \frac{1}{1-\Pi(q)}
= - \frac{1}{\vec{q}\,^2} \left(1 + \Pi(q)\right)
\label{Dq}
\end{equation}
(up to the first correction).

\subsection{Quark and transverse-gluon loops}
\label{S:Quark}

There are several contributions to the Coulomb-gluon self-energy
at one loop.
We begin with the quark-loop contribution $\Pi_q$.
It is Lorentz-invariant,
and is given by the spectral representation
\begin{equation}
\Pi_q(q^2) = \int\limits_0^\infty \frac{\rho_q(s)\,ds}{q^2-s+i0}
\label{Spectral}
\end{equation}
with a positive spectral density
\begin{equation}
\rho_q(s) \geqslant 0\,.
\label{Positive}
\end{equation}

The propagator (and $U_{\vec{q}}$) is a superposition
of the massless propagator and massive ones having various masses,
with positive weights:
\begin{equation}
\begin{split}
U_{\vec{q}} &= - C_F \frac{g_0^2}{\vec{q}\,^2} \left[ 1 -
\int \frac{\rho_q(s)\,ds}{\vec{q}\,^2+s}
+ \cdots \right]\\
&= - C_F g_0^2 \left[
\left( 1 - \int \frac{\rho_q(s)\,ds}{s} \right) \frac{1}{\vec{q}\,^2}
+ \int \frac{\rho_q(s)\,ds}{\vec{q}\,^2+s} + \cdots \right]\,.
\end{split}
\label{Supq}
\end{equation}
Therefore, the potential $U(r)$ is a superposition
of the Coulomb potential and Yukawa ones having various radii,
with positive weights:
\begin{equation}
U(r) = - C_F \frac{g_0^2}{4\pi r} \left[
1 - \int \frac{\rho_q(s)\,ds}{s}
+ \int \rho_q(s)\,e^{-\sqrt{s}\,r}\,ds
+ \cdots \right]\,.
\label{Supr}
\end{equation}
The farther we are from the source,
the more Yukawa potentials switch off,
and the weaker is the interaction.
We have screening.

In QED, this is the only effect.
It seems that screening follows from very general principles:
causality (it allows one to express~(\ref{Spectral})
$\Pi(q^2)$ via its imaginary part $\rho(s)$)
and unitarity (it says that this imaginary part~(\ref{Positive})
is a sum of modulus squared of transition amplitudes
to intermediate states).
The only chance \emph{not} to get screening is to find
some contribution which is not given by the spectral representation.

\begin{figure}[ht]
\begin{center}
\begin{picture}(22,20)
\put(11,10){\makebox(0,0){\includegraphics{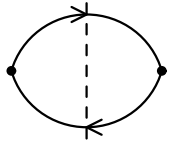}}}
\end{picture}
\end{center}
\caption{Quark loop discontinuity at the cut}
\label{F:qloop}
\end{figure}

Let's complete our calculation.
Using Cutkosky rule (Fig.~\ref{F:qloop}) in the $q$ rest frame,
we see that two integrations are eliminated by two $\delta$-functions,
and
\begin{equation}
\rho_q(s) = T_F n_f \frac{g_0^2 s^{-\varepsilon}}{(4\pi)^{d/2}}
\left( \frac{4}{3} + \mathcal{O}(\varepsilon) \right)
\label{rhoq}
\end{equation}
The ultraviolet divergence of $\Pi_q$ is
\begin{equation*}
\left. \int \frac{\rho_q(s)\,ds}{s+\vec{q}\,^2} \right|_{\text{UV}}
= \frac{4}{3} T_F n_f \frac{g_0^2}{(4\pi)^{d/2}}
\int\limits_{\sim\vec{q}\,^2}^\infty s^{-1-\varepsilon} ds
= \frac{4}{3} T_F n_f \frac{\alpha_s}{4\pi\varepsilon}\,.
\end{equation*}
Keeping only this divergent part, we obtain
\begin{equation}
U_{\vec{q}} = - C_F \frac{g_0^2}{\vec{q}\,^2} \left[ 1
+ \frac{4}{3} T_F n_f \frac{\alpha_s}{4\pi\varepsilon}
+ \cdots \right]\,,
\label{Udq}
\end{equation}
where dots are contributions of other diagrams.

\begin{figure}[ht]
\begin{center}
\begin{picture}(32,20)
\put(16,10){\makebox(0,0){\includegraphics{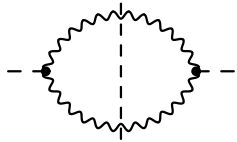}}}
\end{picture}
\end{center}
\caption{Transverse-gluon loop discontinuity at the cut}
\label{F:gt}
\end{figure}

Qualitatively, the transverse-gluon loop
is just like the quark one.
Its contribution in the Coulomb gauge $\Pi_t(q)$
is not Lorentz-invariant,
and the spectral representation is a little more complicated:
\begin{equation}
\Pi_t(q_0^2,\vec{q}\,^2) =
\int \frac{\rho_t(s,\vec{q}\,^2)\,ds}{q^2-s+i0}\,.
\label{Spectral2}
\end{equation}
When calculating the discontinuity by the Cutkosky rule
(Fig.~\ref{F:gt}), we cannot use the $q$ rest frame,
and there is one extra integration
not eliminated by $\delta$-functions.
The general expression for $\rho_t(s,\vec{q}\,^2)$
is rather complicated
(this is the only point in which the Coulomb-gauge derivation
is more complicated than the usual covariant one).

Fortunately, we don't need this general expression.
In order to obtain the ultraviolet divergence of $\Pi_t(q)$,
its limiting form at $s\gg\vec{q}\,^2$ is sufficient.
We can simplify the integrand of our single integral,
and obtain
\begin{equation}
\rho_t(s,\vec{q}\,^2) = C_A \frac{g_0^2 s^{-\varepsilon}}{(4\pi)^{d/2}}
\left( \frac{1}{3} + \mathcal{O}(\vec{q}\,^2/s,\varepsilon) \right)\,.
\end{equation}
The ultraviolet-divergent contribution of this diagram
to $U_{\vec{q}}$ is
\begin{equation}
U_{\vec{q}} = - C_F \frac{g_0^2}{\vec{q}\,^2} \left[ 1
+ \frac{1}{3} C_A \frac{\alpha_s}{4\pi\varepsilon}
+ \cdots \right]\,.
\label{Udt}
\end{equation}

\subsection{Coulomb gluon}
\label{S:Coulomb}

There is one more contribution:
the loop with a Coulomb gluon and a transverse one.
We can understand its sign qualitatively.

\begin{figure}[ht]
\begin{center}
\begin{picture}(40,22)
\put(12,11){\makebox(0,0){\includegraphics{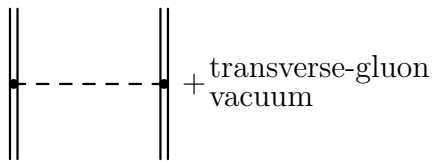}}}
\put(26,11){\makebox(0,0){$+$}}
\put(28,13){\makebox(0,0)[l]{transverse-gluon}}
\put(28,9){\makebox(0,0)[l]{vacuum}}
\end{picture}
\end{center}
\caption{$Q$, $\bar{Q}$, and the transverse-gluon field}
\label{F:QQbarVac}
\end{figure}

Let's consider infinitely heavy quark and antiquark
at a distance $r$ (in the colour-singlet state)
and the transverse-gluon field (Fig.~\ref{F:QQbarVac}).
If we neglect the interaction,
the ground-state energy is just
\begin{equation}
E_0 = U(r)\,,
\label{E0}
\end{equation}
because the vacuum energy of the transverse-gluon field is $0$.

\begin{figure}[ht]
\begin{center}
\begin{picture}(24,22)
\put(12,11){\makebox(0,0){\includegraphics{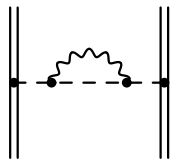}}}
\end{picture}
\end{center}
\caption{Interaction in the second order of perturbation theory}
\label{F:Pert2}
\end{figure}

Now let's take into account their interaction
in the second order of perturbation theory.
Transverse gluons don't interact with infinitely heavy quarks,
they can only couple to Coulomb gluons exchanged between
the quark and the antiquark (Fig.~\ref{F:Pert2}).
The ground-state energy decreases in the second order
of perturbation theory.
This means that the Coulomb attraction becomes stronger ---
antiscreening~\cite{G:77}!

\begin{figure}[ht]
\begin{center}
\begin{picture}(32,12.5)
\put(16,6.25){\makebox(0,0){\includegraphics{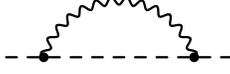}}}
\end{picture}
\end{center}
\caption{Loop with Coulomb and transverse gluons}
\label{F:gc}
\end{figure}

This loop (Fig.~\ref{F:gc}) depends on $\vec{q}$ but not on $q_0$,
because we can always route the external energy $q_0$
via the Coulomb propagator, and it does not depend on energy.
Therefore, this loop has no cut in the $q_0$ complex plane,
and is not given by the spectral representation.
Speaking more formally, we can say that the spectral density is $0$,
and the whole result is given by the subtraction term,
which does not depend on $q_0$ (but depends on $\vec{q}$).

This diagram can be easily calculated.
Only the transverse-gluon propagator depends on $k_0$:
\begin{equation*}
\Pi_c(\vec{q}\,^2) = \int \frac{d^d k}{(2\pi)^d}
\frac{f(\vec{k},\vec{q})}{k^2+i0}\,.
\end{equation*}
The integral in $k_0$ can be taken first:
\begin{equation*}
\int \frac{dk_0}{2\pi} \frac{1}{k_0^2-\vec{k}\,^2+i0}
= - \frac{i}{2 \left(\vec{k}\,^2\right)^{1/2}}\,.
\end{equation*}
We are left with a $(d-1)$-dimensional integral
similar to the usual $d$-dimensional massless loop.
It can be reduced to $\Gamma$-functions via Feynman parametrization.
The result is
\begin{equation}
\Pi_c(\vec{q}\,^2) = C_A
\frac{g_0^2 \left(\vec{q}\,^2\right)^{-\varepsilon}}{(4\pi)^{d/2}}
\left( \frac{4}{\varepsilon} + \mathcal{O}(1) \right)\,.
\label{Pic}
\end{equation}

At last, let's assemble all our findings.
Keeping only ultraviolet-divergent terms in the corrections,
we have the momentum-space potential
\begin{equation}
U_{\vec{q}} = - C_F \frac{g_0^2}{\vec{q}\,^2} \Biggl\{1 +
\frac{g_0^2 (\vec{q}\,^2)^{-\varepsilon}}{(4\pi)^{d/2}}
\biggl[ \left( \left(4 - \frac{1}{3}\right) C_A
- \frac{4}{3} T_F n_f \right) \frac{1}{\varepsilon}
+ \cdots \biggr] \Biggr\}\,.
\label{Utot}
\end{equation}
When expressed via the renormalized $\alpha_s(\mu)$:
\begin{equation}
\frac{g_0^2}{(4\pi)^{d/2}} = \mu^{2\varepsilon} \frac{\alpha_s(\mu)}{4\pi}
Z_\alpha e^{\gamma\varepsilon}\,,\qquad
Z_\alpha = 1 - \beta_0 \frac{\alpha_s}{4\pi\varepsilon}\,,
\label{Renorm}
\end{equation}
this potential must be finite.
Therefore,
\begin{equation}
\beta_0 = \left(4 - \frac{1}{3}\right) C_A
- \frac{4}{3} T_F n_f\,.
\label{beta0}
\end{equation}
The antiscreening term $4$ comes from the Coulomb-gluon loop;
it overweights the screening term $-1/3$ from the transverse-gluon loop.

\subsection{Ward identity}
\label{S;Ward}

Until now, we used the potential between infinitely heavy quark
and antiquark.
Or, if we cut our picture in two halves,
the infinitely-heavy quark -- Coulomb gluon vertex.
This vertex is convenient,
because both the external-leg renormalization
and the vertex corrections vanish (Sect.~\ref{S:Corr}).
However, we can use some other vertex,
for example, the finite-mass quark -- Coulomb gluon vertex,
equally easily.
The tool needed to this end is the Ward identity.
The Coulomb gauge is ghost-free.
Therefore, it is natural to expect that the Ward identities
in this gauge are simple, like in QED,
in contrast to Ward--Slavnov--Taylor identities
in covariant gauges, which are complicated by extra ghost terms.

We shall proceed exactly as in QED.
There, an external photon leg insertion with its polarization 4-vector
parallel to its 4-momentum gives a difference of two propagators;
most terms cancel pairwise.
Now we have a Coulomb gluon, which is polarized along time.
Therefore, let's set its incoming momentum to $q=\omega v$,
where $v=(1,\vec{0})$ is the 4-velocity of the reference frame.
We denote such an external Coulomb gluon by a leg with a black triangle.
A dot near a propagator means that its momentum is shifted by $q$.
It is easy to check the identities
\begin{align}
&\omega \raisebox{-0.4mm}{\includegraphics{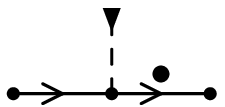}}
= g_0 \raisebox{-0.4mm}{\includegraphics{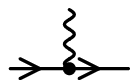}}
\otimes \left[
\raisebox{-0.4mm}{\includegraphics{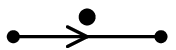}}
- \raisebox{-0.4mm}{\includegraphics{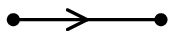}}
\right]\,,
\label{Wardq}\\
&\omega \raisebox{-0.4mm}{\includegraphics{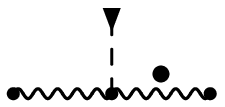}}
= g_0 \raisebox{-3.4mm}{\includegraphics{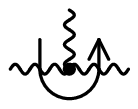}}
\otimes \left[
\raisebox{-0.4mm}{\includegraphics{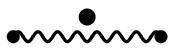}}
- \raisebox{-0.4mm}{\includegraphics{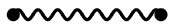}}
\right]\,,
\label{Wardg}\\
&\raisebox{-0.4mm}{\includegraphics{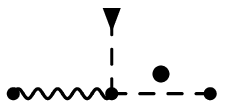}}
= \raisebox{-0.4mm}{\includegraphics{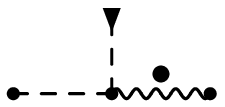}}
= 0\,,
\label{Ward0}
\end{align}
where the right-hand sides are written as
$\text{(colour structure)} \otimes \text{(Lorentz structure)}$,
and the curved arrow in~(\ref{Wardg}) shows the order of indices
in the colour structure $i f^{abc}$.
The last equality~(\ref{Ward0}) is obvious:
let's consider one of these diagrams
with the transverse-gluon line removed;
this object is a vector (has one index),
and depends on two vectors,
$v$ and the transverse-gluon momentum $p$;
and the transverse-gluon propagator is transverse
to both $v$ and $p$.

In covariant gauges, the right-hand side of~(\ref{Wardg})
contains extra terms, where one of the gluon lines becomes
longitudinally polarized.
These terms produce ghost propagators,
thus transforming simple Ward identities
to more complicated Ward--Slavnov--Taylor ones.

Now we are ready to apply these identities
to the one-loop vertex.
Let's consider the QED-like diagram with a transverse gluon:
\begin{equation}
\omega \raisebox{-6.2mm}{\includegraphics{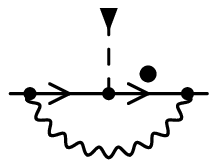}}
= g_0 \raisebox{-4.7mm}{\includegraphics{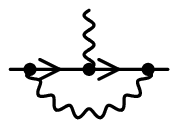}}
\otimes \left[
\raisebox{-2.2mm}{\includegraphics{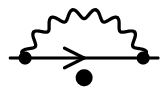}}
- \raisebox{-0.2mm}{\includegraphics{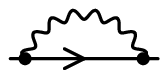}}
\right]\,.
\label{Ward1}
\end{equation}
With a Coulomb gluon,
\begin{equation}
\begin{split}
\omega \raisebox{-6.2mm}{\includegraphics{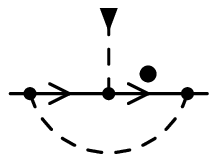}}
&= g_0 \raisebox{-4.7mm}{\includegraphics{qtc.eps}}
\otimes \left[
\raisebox{-2.2mm}{\includegraphics{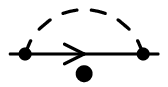}}
- \raisebox{-0.2mm}{\includegraphics{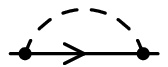}}
\right]\\
&= g_0 \raisebox{-4.7mm}{\includegraphics{qtc.eps}}
\otimes \left[
\raisebox{-0.2mm}{\includegraphics{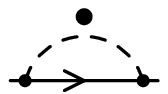}}
- \raisebox{-0.2mm}{\includegraphics{qc3.eps}}
\right] = 0\,,
\end{split}
\label{Ward2}
\end{equation}
because we can route the extra momentum $q$ along the Coulomb-gluon line
instead of the quark one, and the Coulomb propagator does not depend
on an additional momentum $q=\omega v$.
In QED, the only contribution is~(\ref{Ward1})
(and there is no colour factor in its right-hand side):
the difference of the fermion self-energies
with the momenta $p+q$ and $p$.

In QCD, we also have
\begin{equation}
\omega \raisebox{-0.2mm}{\includegraphics{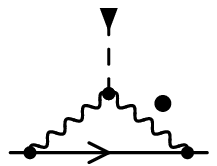}}
= g_0 \left[ \raisebox{-4.7mm}{\includegraphics{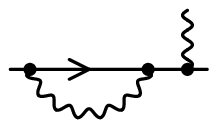}}
- \raisebox{-4.7mm}{\includegraphics{qtc.eps}} \right]
\otimes \left[ \raisebox{-0.2mm}{\includegraphics{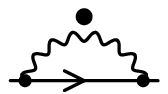}}
- \raisebox{-0.2mm}{\includegraphics{qt3.eps}} \right]\,.
\label{Ward3}
\end{equation}
Here we have used the definition of $i f^{abc}$ via the commutator
in order to re-write the colour structure as a difference:
one term is the same as in~(\ref{Ward1}) (and they cancel);
the second one is the colour structure $C_F$ of the quark self-energy
times that of the elementary vertex $i t^a$.
Two remaining contributions vanish, due to~(\ref{Ward0}):
\begin{equation}
\raisebox{-0.2mm}{\includegraphics{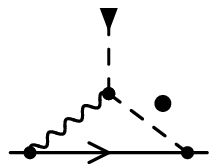}}
= \raisebox{-0.2mm}{\includegraphics{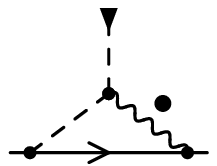}} = 0\,.
\label{Ward4}
\end{equation}
Qualitatively, we can say that the non-abelian charge flows along
both lines in the quark self-energy diagram;
the longitudinal gluon insertion into one of these lines ``measures''
the charge flowing along this line;
in total, the whole quark charge flows,
thus giving the full colour structure of the self-energy $C_F$.

The Ward identity provides a relation
between the quark -- Coulomb gluon vertex
\begin{equation}
\raisebox{-11.7mm}{\begin{picture}(22,25)
\put(11,12.5){\makebox(0,0){\includegraphics{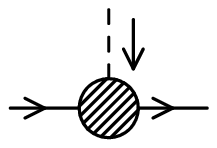}}}
\put(4.5,7){\makebox(0,0)[t]{$p$}}
\put(15.5,15.5){\makebox(0,0)[l]{$q$}}
\end{picture}}
= i g_0 t^a \Gamma(p,q)\,,\qquad
\Gamma(p,q) = \gamma_0 + \Lambda(p,q)
\label{VertDef}
\end{equation}
at $q=\omega v$
and the quark self-energy
\begin{equation}
\raisebox{-4.2mm}{\begin{picture}(22,9)
\put(11,5){\makebox(0,0){\includegraphics{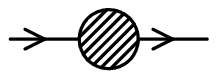}}}
\put(4.5,3){\makebox(0,0)[t]{$p$}}
\end{picture}}
= - i \Sigma(p)
\label{SigmaDef}
\end{equation}
which is related to the propagator:
\begin{equation*}
S(p) = \frac{1}{\rlap/p - m_0 - \Sigma(p)}\,.
\end{equation*}
The Ward identity can be written as
\begin{equation}
\omega \Lambda(p,\omega v) = \Sigma(p) - \Sigma(p+\omega v)
\label{WardLam}
\end{equation}
or
\begin{equation}
\omega \Gamma(p,\omega v) = S^{-1}(p+\omega v) - S^{-1}(p)\,.
\label{WardGam}
\end{equation}
We have proved it at one loop,
but it stays correct at higher loops, too.

Now let's recall how the coupling constant renormalization
\begin{equation*}
g_0 = Z_\alpha^{1/2} g
\end{equation*}
is derived.
When expressed via the renormalized coupling $g$,
the vertex and the propagator should be
\begin{equation*}
\Gamma = Z_\Gamma \Gamma_r\,,\qquad
S = Z_\psi S_r\,,
\end{equation*}
where the renormalized vertex and propagator
are finite at $\varepsilon\to0$.
The scattering amplitude is obtained by multiplying
the proper vertex by the external-leg renormalization factors:
\begin{equation*}
g_0 \Gamma Z_\psi Z_A^{1/2}
= g \Gamma_r Z_\alpha^{1/2} Z_\Gamma Z_\psi Z_A^{1/2}\,,
\end{equation*}
and it must be finite.
Therefore, $Z_\alpha^{1/2} Z_\Gamma Z_\psi Z_A^{1/2}$
must be finite at $\varepsilon\to0$.
But the only minimal renormalization constant
finite at $\varepsilon\to0$ is 1:
\begin{equation}
Z_\alpha = \left( Z_\Gamma Z_\psi \right)^{-2} Z_A^{-1}\,.
\label{Zalpha}
\end{equation}
I.~e., in order to find the coupling-constant renormalization,
one needs the vertex renormalization factor and all
external-leg renormalization factors for this vertex.

The Ward identity makes things simpler.
From~(\ref{WardGam}), $Z_\Gamma Z_\psi$
must be finite at $\varepsilon\to0$, and hence
\begin{equation}
Z_\Gamma Z_\psi = 1\,.
\label{WardZ}
\end{equation}
Therefore,
\begin{equation}
Z_\alpha = Z_A^{-1}\,.
\label{ZalphaWard}
\end{equation}
The coupling constant renormalization is determined by
the Coulomb-gluon propagator renormalization only.
And this is exactly what we studied
in Sect.~\ref{S:Quark}--\ref{S:Coulomb}.

\section{Chromomagnetic properties of vacuum\\
(Nielsen 1981)}
\label{S:Niels}

\subsection{Dielectric and magnetic properties of vacuum}
\label{S:Dielmag}

We shall start from QED,
and then consider non-abelian theory.

We shall use a popular approach to renormalization due to Wilson.
Suppose we have the path integral in momentum space.
We integrate out fields with momenta $p>\Lambda$:
\begin{equation}
\int \prod_p d\phi_p e^{i S} =
\int \prod_{p<\Lambda} d\phi_p e^{i S_\Lambda}\,,
\label{Wilson:path}
\end{equation}
where the effective action is defined by
\begin{equation}
e^{i S_\Lambda} = \int \prod_{p>\Lambda} d\phi_p e^{i S}\,.
\label{Wilson:Seff}
\end{equation}
When we are interested in processes with small momenta $p_i\ll\Lambda$,
$S_\Lambda$ can be expressed via a local Lagrangian.
At the leading order in $1/\Lambda$, it is the standard dimension-4 Lagrangian;
it contains renormalized fields (at the scale $\Lambda$)
and the coupling $g(\Lambda)$.

Now we want to integrate out also fields with momenta
between $\Lambda'$ and $\Lambda$ (where $\Lambda'\ll\Lambda$):
\begin{equation}
e^{i S_{\Lambda'}} = \int \prod_{\Lambda'<p<\Lambda} d\phi_p e^{i S_\Lambda}\,,
\label{Wilson:Renorm}
\end{equation}
and to obtain $g(\Lambda')$.
If we are interested in processes with characteristic momentum $p$,
we should use $\Lambda'$ not too far separated from $p$,
in order to avoid large logarithms.

For example, if we consider interaction between a quark and an antiquark
at a distance $r$,
we should use $\Lambda'\sim1/r$.
Then the Coulomb potential will be simply $-e^2(\Lambda')/r$.
But if we start from the theory with a high cut-off $\Lambda$,
then vacuum modes with momenta between $\Lambda'$ and $\Lambda$
will act as a dielectric medium,
and the potential will be $-e^2(\Lambda)/(\varepsilon r)$:
\begin{equation}
e^2(\Lambda') = \frac{e^2(\Lambda)}{\varepsilon}\,,
\label{Dielmag:Renorm}
\end{equation}
where only modes with momenta between $\Lambda'$ and $\Lambda$
contribute to $\varepsilon$.
If $\varepsilon>1$, we have screening;
if $\varepsilon<1$ --- antiscreening (asymptotic freedom).

In the case of an ordinary matter,
its dielectric and magnetic properties are independent.
But vacuum should be Lorentz-invariant.
Signals should propagate with velocity 1:
\begin{equation}
\varepsilon \mu = 1\,.
\label{Dielmag:Lorentz}
\end{equation}
Therefore, diamagnetic vacuum ($\mu<1$) means screening,
and paramagnetic one ($\mu>1$) --- asymptotic freedom.

When we switch magnetic field $B$ on,
the vacuum energy changes by
\begin{equation}
\Delta E_{\text{vac}} = \left(\mu^{-1} - 1\right) \frac{B^2}{2} V\,.
\end{equation}
We shall show that
\begin{equation}
\Delta E_{\text{vac}} =
- \beta_0 \frac{g^2}{(4\pi)^2} \log\frac{\Lambda^2}{\Lambda^{\prime2}}
\cdot \frac{B^2}{2} V\,.
\label{Dielmag:DEvac}
\end{equation}
This means
\begin{equation}
\mu = \varepsilon^{-1} = 1
+ \beta_0 \frac{g^2}{(4\pi)^2}
\log\frac{\Lambda^2}{\Lambda^{\prime2}}\,,
\label{Dielmag:mu}
\end{equation}
and therefore
\begin{equation}
e^2(\Lambda') = \left[1
+ \beta_0 \frac{e^2}{(4\pi)^2} \log\frac{\Lambda^2}{\Lambda^{\prime2}}
\right]\,e^2(\Lambda)\,,
\end{equation}
i.e., $\beta_0$ is indeed the 1-loop $\beta$-function coefficient.

The vacuum energy of a charged scalar field
(describing particles and antiparticles) is
\begin{equation}
E_{\text{vac}} = 2 \sum_i \frac{\omega_i}{2} = \sum_i \omega_i\,.
\label{Dielmag:Escal}
\end{equation}
For a charged fermion field, it is the energy of the Dirac sea
\begin{equation}
E_{\text{vac}} = - \sum_i \omega_i\,.
\label{Dielmag:Efermi}
\end{equation}
Therefore, in general we can use~(\ref{Dielmag:Escal})
multiplied by $(-1)^{2s}$.

\subsection{Pauli paramagnetism}
\label{S:Pauli}

How the vacuum energy changes when we switch the magnetic field $B$ on?
There are two effects: spin and orbital,
and they can be considered separately.

First we discuss interaction of the spin magnetic moment
with the magnetic field.
Without the field, a massless particle has the energy $\omega=k$.
When the magnetic field is switched on, the energy becomes
\begin{equation}
\omega = \sqrt{k^2 - g_s s_z e B}\,,
\label{Pauli:omega}
\end{equation}
where $g_s$ is the gyromagnetic ratio for our spin-$s$ particle
(we shall discuss this in Sect.~\ref{S:QED} in more detail).

Suppose the magnetic field is along the $z$ axis.
Massless particles only have 2 spin projections $s_z = \pm s$.
The vacuum energy change at the order $B^2$ is
\begin{equation}
\begin{split}
\Delta E_{\text{Pauli}} &{}= (-1)^{2s} \int \frac{V\,d^3 k}{(2\pi)^3}
\left[\sqrt{k^2 + g_s s e B} + \sqrt{k^2 - g_s s e B} - 2 k\right]\\
&{} = - (-1)^{2s} V \frac{(g_s s e B)^2}{4}
\int \frac{d^3 k}{(2\pi)^3} \frac{1}{k^3}
= - (-1)^{2s} V \frac{(g_s s e B)^2}{8 \pi^2} \int \frac{dk}{k}\\
&{} = - 2 (-1)^{2s} (g_s s)^2 \frac{e^2}{(4\pi)^2}
\log\frac{\Lambda^2}{\Lambda^{\prime2}} \cdot \frac{B^2}{2} V\,,
\end{split}
\label{Pauli:DE}
\end{equation}
where only modes with momenta between $\Lambda'$ and $\Lambda$
are included.
Let's stress once more that what we are calculating
is the \emph{vacuum} energy:
there are no particles, only empty modes.

\subsection{Landau levels}
\label{S:LL}

Now let's discuss the effect of magnetic field $B$
on the orbital motion.
In order not to have complications related to spin,
we consider a massless charged scalar field $\varphi$.
Its Lagrangian is
\begin{equation}
L=(D_\mu\varphi)^+ D^\mu\varphi\,.
\label{LL:L}
\end{equation}
For magnetic field $B$ along the $z$ axis,
we can choose the vector potential as $A_y = B x$, $A_x=A_z=0$.
Then the equation of motion is
\begin{equation}
\left[ \nabla^2 - e^2 B^2 x^2 - 2 i e B x \frac{\partial}{\partial y}
+ E^2 \right] \varphi = 0\,.
\label{LL:EOM}
\end{equation}
Its solutions have the form
\begin{equation}
\varphi = e^{i (k_y y + k_z z)} \varphi(x)\,.
\label{LL:sol}
\end{equation}
The equation for $\varphi(x)$ has the same form
as the Schr\"odinger equation for harmonic oscillator:
\begin{equation}
\left[ - \frac{1}{2} \frac{\partial^2}{\partial x^2}
+ \frac{\omega^2}{2} x^2 - E_n \right] \psi_n = 0\,.
\label{LL:osc}
\end{equation}
The oscillator energies are
\begin{equation}
E_n = \omega \left(n + \tfrac{1}{2}\right)\,.
\label{LL:En}
\end{equation}
Comparing~(\ref{LL:EOM}) with~(\ref{LL:osc}),
we see that the energies of our massless particle
in magnetic field $B$ are
\begin{equation}
E^2 = k_z^2 + 2 e B \left(n + \tfrac{1}{2}\right)\,,
\label{LL:LL}
\end{equation}
and the corresponding wave functions are
\begin{equation}
\varphi = e^{i(k_y y + k_z z)} \psi_n\left(x - \frac{k_y}{eB}\right)\,.
\label{LL:phi}
\end{equation}
In other words, $E^2$ consists of discrete Landau levels
of transverse motion plus free motion along the magnetic field.

Each Landau level has a high degree of degeneracy.
In order to find it, let's put our particle
into a large box $V=L_x\times L_y\times L_z$.
Then the allowed longitudinal momenta are
\begin{equation*}
k_z = \frac{2\pi}{L_z} n_z\,;
\end{equation*}
therefore, the number of allowed modes in the interval $d k_z$ is
\begin{equation*}
d n_z = \frac{L_z\,d k_z}{2\pi}\,.
\end{equation*}
Similarly, the allowed values of $k_y$ are
\begin{equation*}
k_y = \frac{2\pi}{L_y} n_y\,.
\end{equation*}
As we see from~(\ref{LL:phi}),
$k_y$ is related to the $x$ coordinate of the center of the Larmor orbit.
It must be inside our box:
\begin{equation*}
\frac{k_y}{eB} \in [0, L_x]\,.
\end{equation*}
Therefore,
\begin{equation*}
n_y \in \left[0, \frac{eB\,L_x L_y}{2\pi}\right]\,.
\end{equation*}
Energy does not depend on $k_y$.
Hence the degeneracy of each Landau level is
\begin{equation}
\frac{eB\,L_x L_y}{2\pi}\,.
\label{LL:degen}
\end{equation}
It is equal to the magnetic flux through our box
($B\,L_x L_y$) measured in flux quanta $2\pi/e$.

\begin{figure}[ht]
\begin{center}
\begin{picture}(52,52)
\put(26,28.5){\makebox(0,0){\includegraphics{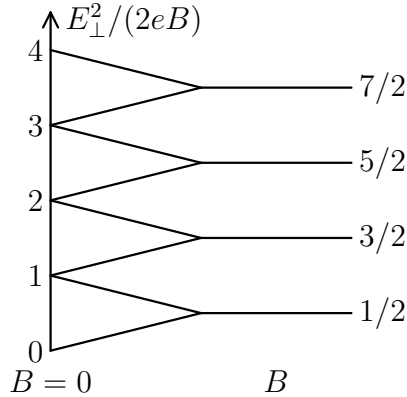}}}
\put(6,2){\makebox(0,0){$B=0$}}
\put(36,2){\makebox(0,0){$B$}}
\put(8,50){\makebox(0,0)[l]{$E_\bot^2/(2eB)$}}
\put(5,6){\makebox(0,0)[r]{0}}
\put(5,16){\makebox(0,0)[r]{1}}
\put(5,26){\makebox(0,0)[r]{2}}
\put(5,36){\makebox(0,0)[r]{3}}
\put(5,46){\makebox(0,0)[r]{4}}
\put(47,11){\makebox(0,0)[l]{$1/2$}}
\put(47,21){\makebox(0,0)[l]{$3/2$}}
\put(47,31){\makebox(0,0)[l]{$5/2$}}
\put(47,41){\makebox(0,0)[l]{$7/2$}}
\end{picture}
\end{center}
\caption{Continuous spectrum at $B=0$ and Landau levels}
\label{F:Landau}
\end{figure}

The spectrum of $E_\bot^2=E^2-k_z^2$ at $B=0$ is continuous.
The number of states in the interval $d E_\bot^2$ is
\begin{equation*}
\frac{L_x L_y}{4\pi} d E_\bot^2\,.
\end{equation*}
When the magnetic field $B$ is switched on,
each interval $\Delta E_\bot^2 = 2eB$ is contracted
into a single Landau level (Fig.~\ref{F:Landau}).
The number of states in each interval is the Landau level
degeneracy~(\ref{LL:degen}).

The vacuum energy of our massless charged scalar field
in the magnetic field $B$ is
\begin{equation}
E_{\text{vac}} = \sum_{n=0}^\infty f\left(n + \tfrac{1}{2}\right)\,,
\label{LL:Evac}
\end{equation}
where
\begin{equation}
f(x) = \frac{eBV}{(2\pi)^2}
\int\limits_{-\infty}^{+\infty} \sqrt{k_z^2 + 2 e B x}\,d k_z\,.
\end{equation}

\subsection{Euler summation formula}
\label{S:Euler}

Integrals are more convenient than sums.
If we have a smooth function $f(x)$
(i.e., its characteristic length is $L\gg1$),
then, obviously,
\begin{equation}
\sum_{n=0}^N f\left(n + \tfrac{1}{2}\right)
\approx \int\limits_0^{N+1} f(x)\,dx\,.
\label{Euler:approx}
\end{equation}
But how to find a correction to this formula?

Let's re-write this integral as a sum of integrals
over unit intervals:
\begin{equation*}
I = \int\limits_0^{N+1} f(x)\,dx =
\sum_{n=0}^N \int\limits_{-1/2}^{1/2} f\left(n + \tfrac{1}{2} + x\right)\,dx\,.
\end{equation*}
The smooth function $f(x)$ can be expanded in Taylor series
in each interval:
\begin{equation*}
I = \sum_{n=0}^N \int\limits_{-1/2}^{1/2}
\left[f\left(n + \tfrac{1}{2}\right)
+ \frac{1}{2} f''\left(n + \tfrac{1}{2}\right) x^2
+ \cdots \right]\,dx
\end{equation*}
(terms with odd powers of $x$ don't contribute).
Calculating the integrals, we get
\begin{equation*}
I = \sum_{n=0}^N f\left(n + \tfrac{1}{2}\right)
+ \frac{1}{24} \sum_{n=0}^N f''\left(n + \tfrac{1}{2}\right)
+ \cdots
\end{equation*}
The second sum in the right-hand side is a small correction
(because $f''\sim f/L^2$);
therefore, we can replace it by an integral
(see~(\ref{Euler:approx})):
\begin{equation*}
I = \sum_{n=0}^N f\left(n + \tfrac{1}{2}\right)
+ \frac{1}{24} \int\limits_0^{N+1} f''(x)\,dx + \cdots
= \sum_{n=0}^N f\left(n + \tfrac{1}{2}\right)
+ \left. \frac{1}{24} f'(x)\right|_0^{N+1} + \cdots
\end{equation*}
Finally, we arrive at the Euler summation formula
\begin{equation}
\sum_{n=0}^N f\left(n + \tfrac{1}{2}\right)
= \int\limits_0^{N+1} f(x)\,dx
- \left. \frac{1}{24} f'(x)\right|_0^{N+1} + \cdots
\label{Euler:Sum}
\end{equation}
The correction here is of order $1/L^2$.
It is easy to find a few more corrections,
if desired.

\subsection{Landau diamagnetism}
\label{S:Landau}

The vacuum energy in magnetic field $B$~(\ref{LL:Evac})
can be re-written using the Euler formula~(\ref{Euler:Sum}) as
\begin{equation}
E_{\text{vac}} = \int\limits_0^\infty f(x)
- \left. \frac{1}{24} f'(x) \right|_0^\infty\,.
\label{Landau:Evac}
\end{equation}
The integral here is the vacuum energy at $B=0$
(when the spectrum of $E_\bot^2$ is continuous, Fig.~\ref{F:Landau}).
The shift of the vacuum energy due to the magnetic field is
\begin{equation}
\Delta E_{\text{Landau}} = \frac{1}{24} f'(0)
= \frac{e^2 B^2 V}{48 \pi^2} \int \frac{dk}{k}
= \frac{1}{3} \frac{e^2}{(4\pi)^2} \log\frac{\Lambda^2}{\Lambda^{\prime2}}
\cdot \frac{B^2}{2} V\,,
\label{Landau:DE}
\end{equation}
where only modes with momenta between $\Lambda'$ and $\Lambda$
are included%
\footnote{Many subtleties have been swept under the carpet
in this derivation sketch.
A somewhat more accurate derivation~\cite{N:81} yields the same result.}.

\subsection{The QED result}
\label{S:QED}

The full $\Delta E_{\text{vac}}$ in QED
is the sum of the spin contribution~(\ref{Pauli:DE})
and the orbital one~(\ref{Landau:DE}):
\begin{equation}
\Delta E_{\text{vac}} = - \beta_0 \frac{e^2}{(4\pi)^2}
\log\frac{\Lambda^2}{\Lambda^{\prime2}} \cdot \frac{B^2}{2} V\,,
\label{QED:DE}
\end{equation}
where
\begin{equation}
\beta_0 = 2 \sum_s (-1)^{2s} \left[ (g_s s)^2 - \frac{n_s}{6} \right]
\label{QED:beta0}
\end{equation}
is the sum over all charged fields,
and $n_s$ is the number of polarization states: $n_0=1$, $n_{s\neq0}=2$.

Let's demonstrate that a Dirac particle (e.g., electron)
has gyromagnetic ratio $g_{1/2}=2$.
For a massless particle, the Dirac equation is
\begin{equation}
\D\psi = 0\,,\qquad
D_\mu = \partial_\mu - i e A_\mu\,.
\label{QED:Dirac}
\end{equation}
Let's multiply it by $\D$:
\begin{equation*}
\D^2 \psi = 0\,,\qquad
\D^2 = \partial^2 - e^2 A^2 - 2 i e A^\mu \partial_\mu
- i e \gamma^\mu \gamma^\nu \partial_\mu A_\nu
\end{equation*}
(in the last term, $\partial_\mu$ acts only on $A_\nu$).
If we suppose that $\partial\cdot A=0$,
\begin{equation*}
\D^2 = D^2 - \frac{ie}{4} F_{\mu\nu} [\gamma^\mu,\gamma^\nu]\,.
\end{equation*}
We choose $A^\mu=(0,0,B x^1,0)$, then $F_{12}=-F_{21}=-B$ and
\begin{equation*}
\D^2 = D^2 + i e B \gamma^1 \gamma^2 = D^2 + 2 e B s_z\,.
\end{equation*}
The equation of motion in magnetic field $B$
(directed along $z$) becomes
\begin{equation}
\left[ \nabla^2 - e^2 B^2 x^2 - 2 i e B x \frac{\partial}{\partial y}
+ 2 e B s_z + E^2 \right] \psi = 0\,.
\label{QED:EOM}
\end{equation}
Its energy spectrum is
\begin{equation}
E^2 = k_z^2 + 2 e B \left(n + \tfrac{1}{2}\right) - 2 e B s_z
\label{QED:E}
\end{equation}
(see Sect.~\ref{S:LL})%
\footnote{Strictly speaking, we should consider
spin and orbital effects together, using~(\ref{QED:E}).
It is easy to see that in the order $B^2$ they can be treated separately.}.
Comparing this with~(\ref{Pauli:omega}), we see that
\begin{equation}
g_{1/2} = 2\,.
\label{QED:g}
\end{equation}

So, the electron contribution to $\beta_0$~(\ref{QED:beta0}) is $-4/3$;
if a charged scalar particle exists, it contributes $-1/3$.
Note that for $s=1/2$ the spin effect overweights the orbital one.
However, due to the factor $(-1)^{2s}$,
the spin effect leads to diamagnetism,
and the orbital one to paramagnetism.
This is because here we are interested in the Dirac see.
In physics of metals, we are interested in positive-energy electrons
(below the Fermi surface),
and the spin effect gives Pauli paramagnetism,
while the orbital one --- Landau diamagnetism.

\subsection{The QCD result}
\label{S:QCD}

Now it is easy to obtain $\beta_0$ in QCD.
We have chromomagnetic field instead of magnetic.
Let's choose its colour orientation along an axis $a_0$
such that $t^{a_0}$ is diagonal
(for the $SU(3)$ colour group with the standard choice of the generators $t^a$,
$t^8$ is diagonal, and we choose $a_0=8$).

The quark contribution follows from the electron one in QED.
The contribution to $\beta_0$ is proportional to the charge squared.
The sum of squares of colour ``charges'' of a quark is $\Tr t^{a_0} t^{a_0}$
(no summation).
Recalling
\begin{equation*}
\Tr t^a t^b = T_F \delta^{ab}\,,
\end{equation*}
we arrive at the contribution
\begin{equation}
- \left(1 - \tfrac{1}{3}\right) 2 T_F n_f
\label{QCD:quark}
\end{equation}
of $n_f$ quark flavours.
If there were scalar quarks, each flavour would contribute
\begin{equation*}
- \frac{1}{3} 2 T_F\,.
\end{equation*}

What about gluons?
First of all, we have to find their gyromagnetic ratio $g_1$.
We shall do this for the $SU(2)$ colour group,
because calculations are simpler in this case;
the result will be valid for other colour groups, too.
Let's consider the $SU(2)$ Yang--Mills equation
\begin{equation}
D^\nu G^a_{\mu\nu}
= \left( \partial^\nu \delta^{ab} + g \varepsilon^{acb} A^{c\nu} \right)
G^b_{\mu\nu} = 0\,.
\label{QCD:YM}
\end{equation}
The external field is $A^3_\mu$.
We linearize in the small components $A^{1,2}_\mu$.
For $A^-_\mu=A^1_\mu-i A^2_\mu$ we get
\begin{equation*}
D^\nu G^-_{\mu\nu} + i g G^3_{\mu\nu} A^{-\nu} = 0\,,
\end{equation*}
where
\begin{equation*}
G^-_{\mu\nu} = D_\mu A^-_\nu - D_\nu A^-_\mu\,,\qquad
D_\mu = \partial_\mu - i g A^3_\mu\,.
\end{equation*}
In the $D^\mu A^-_\mu=0$ gauge, the equation of motion becomes
\begin{equation*}
D^2 A^-_\mu - 2 i g G^3_{\mu\nu} A^{-\nu} = 0
\end{equation*}
(we have used $[D_\mu,D_\nu]=-i g G^3_{\mu\nu}$).
Our external field is oriented along $z$ in space
and along 3 in colour: $G^3_{12} = -G^3_{21} = -B$.
Using $A^{-2} = i s_z A^{-1}$ ($s_z=\pm1$), we finally obtain
\begin{equation}
\left[ D^2 + 2 i g B s_z \right] A^{-1} = 0\,.
\label{QCD:EOM}
\end{equation}
This equation looks exactly the same as~(\ref{QED:EOM}),
and hence $g_1=2$.

Gluons with colour $a_1$ such that $t^{a_1}$ is diagonal
don't interact with our chromomagnetic field
(for the standard $SU(3)$ generators, $a_1=3$).
All other gluons can be arranged into pairs with positive and negative
``colour charges'' (particles and antiparticles).
The sum of their squares (both signs!) is $C_A$:
in the adjoint representation
\begin{equation*}
\Tr t^a t^b = C_A \delta^{ab}\,,
\end{equation*}
i.e.\ we have to replace $2 e^2 \to C_A g^2$ in QED results.

Finally, we arrive at
\begin{equation}
\beta_0 = \left(4 - \tfrac{1}{3}\right) C_A
- \left(1 - \tfrac{1}{3}\right) 2 T_F n_f\,.
\label{QCD:beta0}
\end{equation}
Pauli paramagnetism of the gluon vacuum $(g_1\cdot1)^2=4$ is stronger than
its Landau diamagnetism $-1/3$.
This leads to antiscreening (asymptotic freedom).

\section*{Acknowledgments}

I am grateful to I.B.~Khriplovich for discussions of~\cite{Kh:69}.


\begin{thebibliography}{99}

\bibitem{VT:65}
V.S.~Vanyashin, M.V.~Terentev,
ZhETF \textbf{48} (1965) 565\\
{}[Sov.\ Phys.\ JETP \textbf{21} (1965) 375]

\bibitem{Kh:69}
I.B.~Khriplovich,
Yad.\ Phys.\ \textbf{10} (1969) 409
[Sov.\ J.\ Nucl.\ Phys.\ \textbf{10} (1970) 235]

\bibitem{tH:72}
G.~'t~Hooft, Marseilles meeting on field theory (1972), unpublished;\\
Nucl.\ Phys.\ \textbf{B62} (1973) 444

\bibitem{GW:73}
D.J.~Gross, F.~Wilczek,
Phys.\ Rev.\ Lett.\ \textbf{30} (1973) 1343

\bibitem{P:73}
H.D.~Politzer,
Phys.\ Rev.\ Lett.\ \textbf{30} (1973) 1346

\bibitem{Sh:01}
M.~Shifman,
in \textit{At the frontiers of particle physics},
World Scientific (2001), v.~1, p.~126

\bibitem{tH:01}
G.~'t~Hooft,
in \textit{The creation of quantum chromodynamics and the effective energy},
World Scientific (2001), p.~9

\bibitem{W:96}
F.~Wilczek,
hep-th/9609099

\bibitem{Nobel}
D.J.~Gross, Rev.\ Mod.\ Phys.\ \textbf{77} (2005) 837;\\
H.D.~Politzer, Rev.\ Mod.\ Phys.\ \textbf{77} (2005) 851;\\
F.~Wilczek, Rev.\ Mod.\ Phys.\ \textbf{77} (2005) 857

\bibitem{W2}
S.~Weinberg,
\textit{The quantum theory of fields}, v.~2,
Cambridge University Press (1996)

\bibitem{G:07}
A.G.~Grozin,
\textit{Lectures on QED and QCD: practical calculation and renormalization
of one- and multi-loop Feynman diagrams},
World Scientific (2007)

\bibitem{G:04}
A.G.~Grozin,
\textit{Heavy quark effective theory},
Springer (2004)

\bibitem{A:81}
L.F.~Abbott,
Nucl.\ Phys.\ \textbf{B185} (1981) 189;
Acta Phys.\ Polon.\ \textbf{B13} (1982) 33

\bibitem{N:81}
N.K.~Nielsen,
Amer.\ J.\ Phys.\ \textbf{49} (1981) 1171

\bibitem{NRQCD}
N.~Brambilla, A.~Pineda, J.~Soto, A.~Vairo,
Rev.\ Mod.\ Phys.\ \textbf{77} (2005) 1423

\bibitem{BB:94}
M.~Beneke, V.M.~Braun,
Nucl.\ Phys.\ \textbf{B426} (1994) 301

\bibitem{G:77}
V.N.~Gribov,
in \textit{12 Winter School of the Leningrad Nuclear Physics Inst.} (1977);
translation in \textit{The Gribov theory of quark confinement},
ed.\ J.~Nyiri, World Scientific (2001), p.~24

\end{thebibliography}
\end{document}